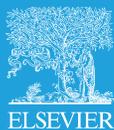

*HIGHLIGHTED PAPER*

# Topological iron silicide with H* intermediate modulated surface for efficient electrocatalytic hydrogenation of nitrobenzene in neutral medium

Yuchen Wang [1],[#], Yaoyu Liu [1],[#], Zhiyue Zhao [1], Zhikeng Zheng [1], Alina M. Balu [2],[*], Rafael Luque [3],[4], Kai Yan [1],[*]

[1] Guangdong Provincial Key Laboratory of Environmental Pollution Control and Remediation Technology, School of Environmental Science and Engineering, Sun Yat-sen University, Guangzhou 510275, China
[2] Departamento de Química Orgánica, Universidad de Córdoba, Campus Universitario de Rabanales, Edificio Marie Curie (C3), Córdoba E-14014, Spain
[3] Center for Refining and Advanced Chemicals, King Fahd University of Petroleum and Minerals, Dhahran 31261, Saudi Arabia
[4] Universidad ECOTEC, Km 13.5 Samborondón, Samborondón EC092302, Ecuador

Electrocatalytic hydrogenation of nitrobenzene (Ph-NO$_2$) reaction (EHNR) has been considered as a potential alternative to the traditional thermocatalytic process in the production of high-value aniline (Ph-NH$_2$). However, due to the absence of robust catalyst and low surface H* coverage, the EHNR faces the challenges of undesired performance and indetermined mechanism. Herein, we construct a type of noble-metal free topological FeSi (M-FeSi) materials through a solvent-free microwave strategy for efficient EHNR in neutral medium. Impressively, benefiting from abundant active H* intermediates on the surface of M-FeSi catalyst, the topological M-FeSi catalyst exhibits 99.7% conversion of Ph-NO$_2$ and 93.8% yield of Ph-NH$_2$ after 200 C in neutral medium, which are superior to previous candidates and FeSi catalyst synthesized via the traditional arc-melting method under same conditions. Besides, theoretical calculations validate that high surface H* coverage over M-FeSi catalyst is conducive to switching the rate-determining step from Ph-NO$_2$* → Ph-NO* to Ph-NO* → Ph-NHOH*, and thus decreasing the total energy barrier of electrocatalytic Ph-NH$_2$ production.

## Introduction

Aniline (Ph-NH$_2$) is an indispensable high-value chemical stock in various industries, such as pigments and drugs [1]. Currently, the state-of-art technique for Ph-NH$_2$ production is thermocatalytic hydrogenation of nitrobenzene (Ph-NO$_2$) by employing a number of noble-metal catalysts (e.g., Pt, Pd) or reductants (e.g., hydrogen gas) [2]. Albeit those processes have exhibited high yield and desired selectivity, the accompanied elevated temperature and high pressure may cause potential safety risks [3]. Therefore, it is essential to develop a robust and durable catalyst in a greener manner for the efficient hydrogenation of Ph-NO$_2$ toward Ph-NH$_2$ at the ambient environment.

Electrocatalytic hydrogenation of Ph-NO$_2$ reaction (EHNR) has recently attracted tremendous attention for its sustainable process, high efficiency, and benign reaction condition [4]. EHNR is greatly affected by the pH value of electrolyte [5], since the overall reactivity of EHNR is closely related to H* intermediates on the surface of a catalyst, which are in situ formed through electrocatalytic reduction of protons or water molecules from electrolyte [6]. On one hand, in low pH environment, the competitive hydrogen evolution reaction (HER) consumes plenty of H* intermediates to generate hydrogen gas, leading to reduced electrocatalytic hydrogenation activity [7]. On the other hand, the existence of abundant hydroxide ions in high pH environ-

   

ment could decrease the generation rate of H* intermediates in the Volmer reaction (H$_2$O + e$^-$ → H* + OH$^-$), which also suppresses electrocatalytic hydrogenation activity. In this context, the rational fabrication of a robust and durable catalyst is vital to drive EHNR in different media. Zhao et al. employed a plasma-aided exfoliation strategy to fabricate sulfur-deficient Co$_3$S$_4$ ultrathin nanosheets, which offered Ph-NO$_2$ conversion of 98% and Ph-NH$_2$ selectivity of 99% in alkaline electrolyte [8]. Very recently, Co$_9$S$_8$/Ni$_3$S$_2$ heterojunctions were constructed by a self-template method and also reported as high-performance electrocatalysts for EHNR [9]. Benefiting from the directional electron transfer between counterparts of the heterojunction, the enhanced co-adsorption of water and Ph-NO$_2$ molecules at the interface resulted in optimized Ph-NH$_2$ selectivity of 96% with Faradaic efficiency (FE) of 95.3% in alkaline environment. Up to date, albeit certain progress have been made for alkaline EHNR, few catalytsts are designed and explored for this process in neutral medium.

A high interest has emerged on transition metal monosilicides (TMSi) in recent years for their attractive physical properties, such as molecular dissociation [10], double-Weyl phonons [11], helicoid-arc quantum states [12], chiral fermions [13]. In particular, FeSi, CoSi, MnSi, ReSi, and RuSi have been predicted to exhibit novel topological points, and RhSi presents large topological Fermi arcs [11,13–14]. These topological TMSi have recently exhibited a remarkable potential as electrocatalysts with high electrocatalytic HER activity in theoretical and experimental aspects [15]. Notably, our group have demonstrated that the boosted HER activity of TMSi in acidic electrolyte were attributed to the enhanced H* adsorption ability of TM-Si bond [15a]. Inspired by this work, TMSi is believed to potentially act as efficient EHNR with the aid of sufficient H* intermediates in neutral medium.

Herein, we have successfully designed topological FeSi (M-FeSi) electrocatalysts through a solvent-free one-step microwave strategy with prospects on a significant enhancement of the EHNR process as compared to their counterparts in neutral medium (Scheme 1). The microstructure of the controllably synthesized M-FeSi was firstly investigated by comprehensive characterization techniques (e.g., high-resolution transmission electron microscopy and X-ray absorption fine structure spectroscopy). Subsequently, the EHNR performance of the M-FeSi catalyst were studied in details and key experimental parameters (e.g., the reaction electrolyte and applied potential) were adjusted to optimize the EHNR activity. It is noteworthy that the importance of active H* intermediates generated on the surface of M-FeSi catalyst during neutral EHNR process was identified and elucidated by advanced operando characterization tools (e.g., X-band electron paramagnetic resonance, rotating ring-disk electrode technique, and electrochemical impedance spectroscopy) and theoretical calculations. Our finding could offer a promising strategy for rational design of topological TMSi catalysts for sustainable electrocatalytic hydrogenation reaction.

## Results and discussion

M-FeSi were synthesized in a one-step solventless microwave approach. As depicted in Fig. 1**a**, Fe and Si pure powders with the molar ratio of 1:1 were initially ground together and then melted to form M-FeSi under suitable dielectric strength, which was controlled by a microwave generator. The detailed synthetic process has been included in the Supporting Information.

X-ray diffraction (XRD) pattern firstly validated the successful preparation of topological M-FeSi with classical cubic structure (Fig. 1b) [15a,16]. According to Bragg's equation, the lattice constant can be estimated as 4.488 Å, which is nearly identical to the theoretical value (4.467 Å). Subsequently, morphological characterizations including scanning electron microscopy (SEM) and transmission electron microscopy (TEM) images (Fig. 1c, d, and Fig. S1, Supporting Information) presented that

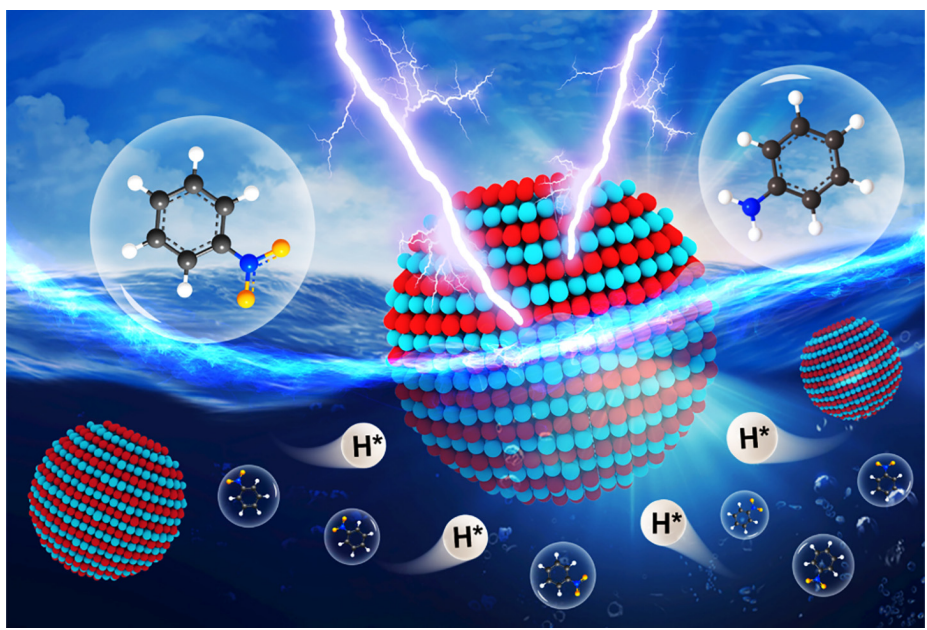

**SCHEME 1**

Schematic illustration of M-FeSi electrocatalyst with the modulated H* intermediates for accelerating EHNR.

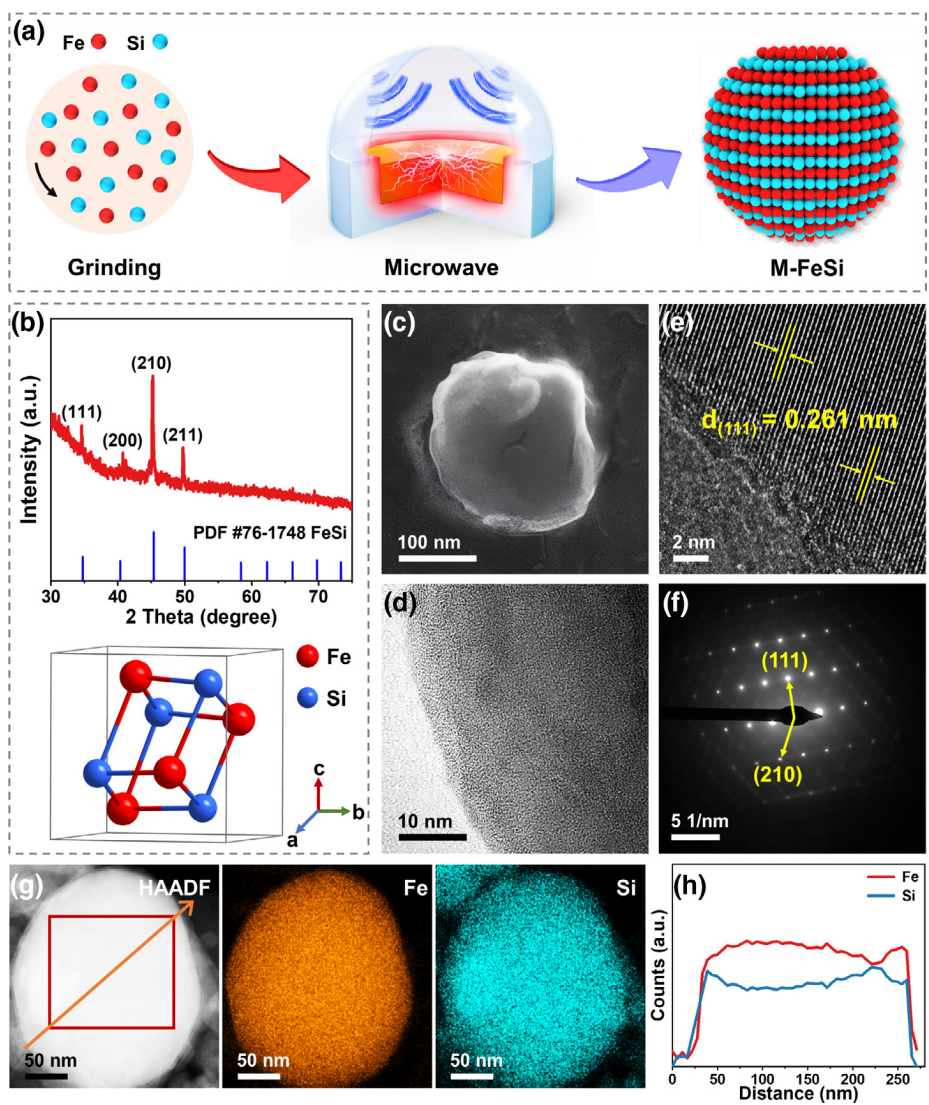

FIGURE 1

a) Schematic illustration of the synthesis of M-FeSi using the microwave method. b) XRD pattern and crystal structure of M-FeSi. c) SEM, d) TEM, e) HR-TEM, and f) SAED images of M-FeSi. g) HAADF-STEM image, corresponding spatial elemental mapping, and h) line scan profiles of M-FeSi.

as-synthesized M-FeSi was composed of ellipsoidal particles. Meanwhile, the corresponding high-resolution TEM (HR-TEM) and selected-area electron diffraction (SAED) images (Fig. 1e, f) further confirmed the existence of FeSi phase, consistent with XRD results. In addition, the high-angle annular dark-field scanning transmission electron microscopy (HAADF-STEM) and corresponding element mapping images (Fig. 1g, h and Fig. S2, Supporting Information) revealed the homogeneous distribution of Fe and Si elements along ellipsoidal particles.

X-ray absorption fine structure spectroscopy (XAFS) was performed to decipher the electronic structure of M-FeSi. As displayed in Fig. 2a, the X-ray absorption near edge structure (XANES) spectrum of M-FeSi was prone to be overlapped with that of reference Fe foil, suggesting that the oxidation state of Fe element in M-FeSi remains almost zero. Fig. 2b presents the corresponding Fourier-transformed extended X-ray absorption fine structure (FT-EXAFS) spectra. Compared with reference samples of Fe foil and $Fe_2O_3$, the main peak of M-FeSi at 1.90 Å can be assigned to the Fe-Si bond, locating between the Fe-O bond (1.44 Å) and the Fe-Fe bond (2.21 Å). Furthermore, the coordination environment of M-FeSi was intuitively visualized in the wavelet-transformed EXAFS (WT-EXAFS) spectra. In Fig. 2c, the area of Fe-Si bond was centered at approximately 5.85 Å$^{-1}$, which was distinctly distinguished from that of Fe-O and Fe-Fe bonds. Therefore, combining EXAFS results in both R space and K space, it is clearly denoted that pure phase of topological M-FeSi is formed without the interference of heteroatoms (C, N, O, etc.) or the aggregation of Fe [17].

To investigate the electrocatalytic hydrogenation activity and corresponding product analysis of M-FeSi, the electrochemical characterizations were carried out using a H-shaped electrochemical cell with Nafion separator (Fig. S3, Supporting Information). In this three-electrode system, the working electrode was prepared through a hydrothermal method to realize in situ growth of M-FeSi powders on the carbon fiber paper (CFP), which kept the crystal structure of M-FeSi unchanged (Fig. S4, Supporting Information). Fig. 3a-c display linear sweep voltammetry (LSV) curves of M-FeSi electrocatalyst in wide pH range with and with-

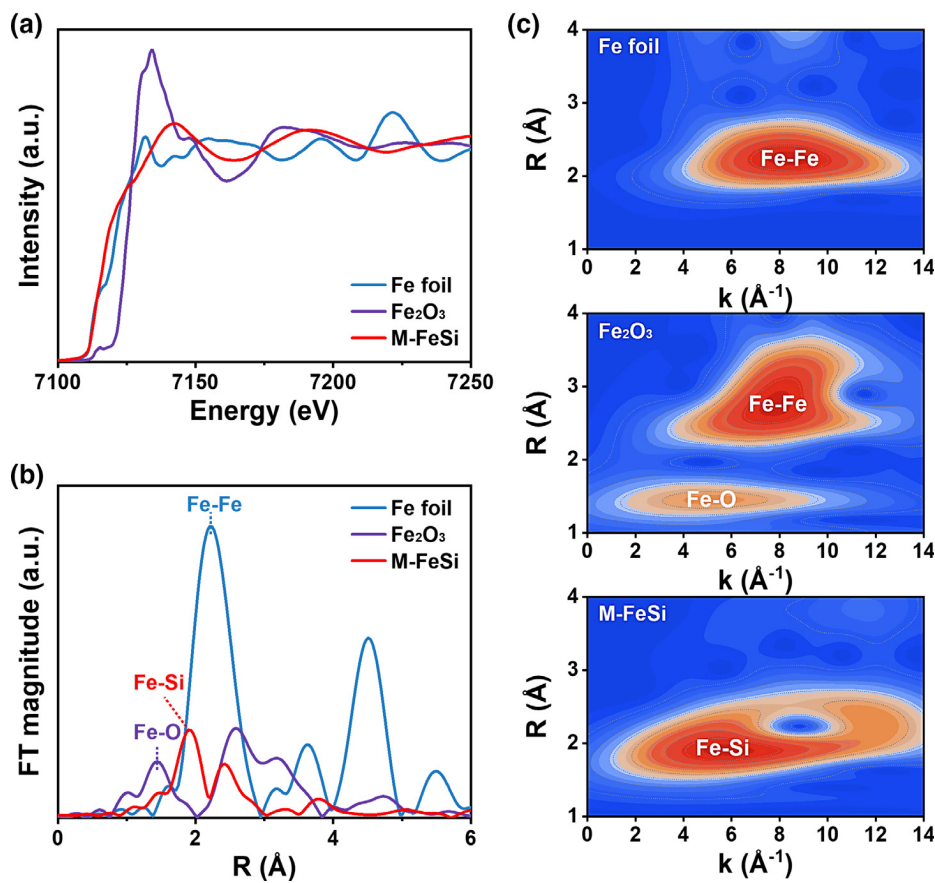

**FIGURE 2**

a) XANES spectra, b) FT-EXAFS spectra, and c) WT-EXAFS spectra of Fe foil, $Fe_2O_3$, and M-FeSi, respectively.

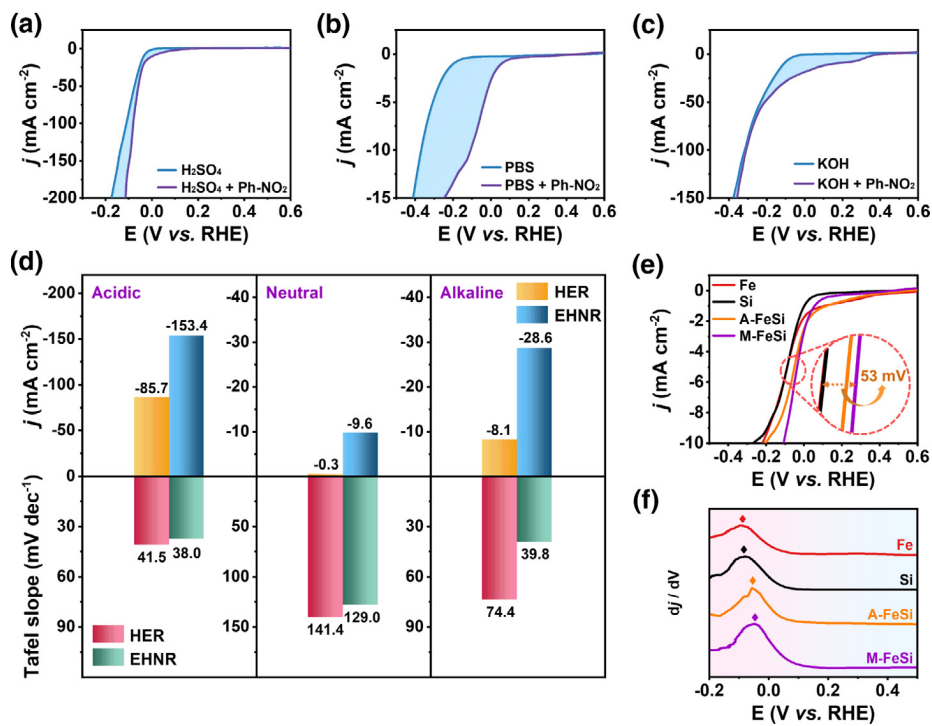

**FIGURE 3**

LSV curves of M-FeSi electrocatalyst in a) 0.5 M $H_2SO_4$, b) 0.01 M PBS, and c) 1 M KOH with/without 10 mM Ph-$NO_2$. All LSV curves were corrected with iR compensation. d) The HER and EHNR performance comparison of M-FeSi (current density at −0.1 V vs. RHE, top; Tafel plot, bottom) in acidic, neutral, and alkaline media, respectively. e) LSV curves and (f) corresponding half-wave potential of Fe, Si, A-FeSi, and M-FeSi electrocatalysts in 0.01 M PBS with the presence of 10 mM Ph-$NO_2$.

out Ph-NO$_2$. All LSV curves in this work were shown with iR-correction based on Nyquist plots (Fig. S5, Supporting Information). After adding 10 mM Ph-NO$_2$, all LSV curves in three media exhibited enhanced current density, indicating that EHNR is occurred prior to hydrogen evolution. Concretely, M-FeSi achieved the largest current density of 153.4 mA cm$^{-2}$ at −0.1 V vs. RHE in 0.5 M sulfuric acid (H$_2$SO$_4$) with 10 mM Ph-NO$_2$ (Fig. 3d). Subsequently, the kinetics of EHNR and HER over M-FeSi electrocatalyst were assessed by Tafel slope (Fig. 3d and Fig. S6, Supporting Information). The Tafel slope values for hydrogenating 10 mM Ph-NO$_2$ were derived as 38.0, 129.0, and 39.8 mV dec$^{-1}$ in 0.5 M H$_2$SO$_4$, 0.01 M phosphate buffer solution (PBS), and 1 M potassium hydroxide (KOH), respectively, which were lower than that of HER. The pronounced reduction of Tafel slope manifests that EHNR is kinetically more favorable than HER over M-FeSi.

Since H* intermediates in neutral electrolyte could be only derived from water dissociation, M-FeSi, FeSi synthesized using arc-melting method (A-FeSi), counterpart Fe, and Si were tested in 0.01 M PBS with 10 mM Ph-NO$_2$ to assess their catalytic performance of EHNR in H*-deficient environment. In comparison with M-FeSi, A-FeSi possessed a larger bulky structure with micrometer size (Fig. S7, Supporting Information). As presented in Fig. 3e, the driving potential of M-FeSi electrocatalyst at a current density of 5 mA cm$^{-2}$ was 53, 53, and 14 mV lower than that of Fe, Si, and A-FeSi, respectively. The possible reason is that Fe-Si bond formed under microwave radiation is favorable for the in situ generation of active H* intermediates, which are beneficial for EHNR [8a]. Correspondingly, the lowest half-wave potential (Fig. 3f) and the largest turnover frequency (TOF) value (Fig. S8, Supporting Information) further confirmed the superior electrocatalytic performance for EHNR [5]. Besides, the long-term stability of M-FeSi was evaluated via multiple chronoamperometric cycles at the potential of −0.4 V vs. RHE (Fig. S9, Supporting Information). For each cycle, the current density dropped to around 7 mA cm$^{-2}$ and recovered back to the initial value after adding fresh electrolyte. In combination with unchanged LSV curves (Fig. S10, Supporting Information), XRD patterns (Fig. S11, Supporting Information) and SEM images of M-FeSi (Fig. S12, Supporting Information) after cycles, topological M-FeSi electrocatalyst was demonstrated to possess an excellent durability, showing a great significance for practical utilization.

The potential-dependent selectivity of EHNR over M-FeSi was studied at different potentials from −0.2 to −0.6 V vs. RHE by applying chronoamperometric measurements. High-performance liquid chromatography (HPLC) was used to quantify standard substances and products (Figs. S13-S15, Supporting Information). As shown in Fig. S16a (Supporting Information), azoxybenzene (Ph-N=NO-Ph) and p-aminophenol (OH-Ph-NH$_2$) were observed at the low potential of −0.2 V vs. RHE in acidic medium. With the increasing potential, OH-Ph-NH$_2$ became dominant with very high selectivity, indicating that Ph-NO$_2$ is not likely reduced to Ph-NH$_2$ in the acidic medium. Differently, the selectivity of EHNR in neutral medium was irrelevant to the applied reduction potential and Ph-NH$_2$ was the dominant hydrogenation product with approximately 100% selectivity (Fig. S16b, Supporting Information). In the case of alkaline condition, the main hydrogenation product was also found as Ph-NH$_2$ (Fig. S16c, Supporting Information). Nevertheless, the appearance of phenylhydroxylamine (Ph-NHOH) and azobenzene (Ph-N=N-Ph) within the investigated potential range denoted that the reaction pathways of Ph-NH$_2$ generation are not identical in neutral and alkaline medium. Considering the high selectivity of Ph-NH$_2$, the optimal potential was chosen as −0.4 V vs. RHE and the corresponding charge-dependent yields of hydrogenation product are presented in Fig. S17 (Supporting Information). Compared with acid and alkaline conditions, neutral EHNR achieved higher Ph-NH$_2$ yield after passing a fixed charge of 200 C. Moreover, under the same condition (0.01 M PBS, 200 C), M-FeSi realized outperformed Ph-NO$_2$ conversion (99.7%), FE (81.5%) and Ph-NH$_2$ yield (93.8%) over Fe, Si, and A-FeSi (Fig. S18, Supporting Information). These outstanding EHNR performance over M-FeSi are comparable or superior to that of recent reported electrocatalysts in Ph-NH$_2$ production (Table S1, Supporting Information).

Based on above product analysis, possible pathways for EHNR over M-FeSi electrocatalyst can be proposed and plotted in Fig. S19 (Supporting Information). The pathways are close related to the hydrogenation environment. Specifically, EHNR is inclined to produce OH-Ph-NH$_2$ in acidic medium (Path I, purple), which stem from the Bamberger rearrangement of Ph-NHOH [18]. Afterward, direct electrocatalytic hydrogenation of Ph-NO$_2$ toward Ph-NH$_2$ without any side reaction is easily proceeded in neutral medium (Path II, red). Alternatively, in alkaline medium, extra condensation reaction between intermediates nitrosobenzene (Ph-NO) and Ph-NHOH is more likely occurred to form Ph-N=NO-Ph, which can be further reduced to Ph-NH$_2$ (Path III, green) [19]. Therefore, combining superior electrochemical performance and efficient Ph-NH$_2$ generation, it is unambiguous to conclude that the neutral medium is the optimal environment for EHNR over M-FeSi electrocatalyst.

Attention was then directed to the mechanistic study of EHNR over M-FeSi electrocatalyst in neutral medium. In general, electrocatalytic hydrogenation process proceeds through either direct reaction on the active sites of electrocatalysts or indirect electron transfer with the assistance of active reductive species, such as H* [2b]. The type of EHNR mechanism over M-FeSi was determined by CV curves with different starting potentials (from −0.07 to −0.37 V vs. RHE) and a fixed ending potential at 1.02 V vs. RHE. In pure PBS solution (Fig. 4a), inconspicuous oxidation peaks at 0.318–0.346 V vs. RHE were observed and referred to H* adsorption [20]. With the addition of Ph-NO$_2$, strong peaks emerged in potential ranges of 0.020–0.024, 0.233–0.236, and 0.330–0.360 V vs. RHE (Fig. 4b), which could be attributed to the adsorption/desorption of H* intermediates [20]. The appearance of new peaks and the amplification of existing peak implied the formation of abundant H* intermediates on the surface of electrode, suggesting that EHNR occurs through indirect reaction mechanism over M-FeSi catalyst.

To in-depth investigate the relationship between H* intermediates and EHNR activity on M-FeSi, quasi in situ electron paramagnetic resonance (EPR) measurements were carried out using 5,5-dimethyl-1-pyrroline N-oxide (DMPO) as spin-trapping reagents. As illustrated in Fig. 4c, after applying −0.4 V vs. RHE on M-FeSi for 10 min, the occurrence of nine characteristic peaks of DMPO-H adduct signified the existence of active H* intermediates [21], which is in accordance with CV results. More importantly, the addition of Ph-NO$_2$ led to the enhanced EPR signal, indicating that the formation of H* intermediates is accelerated during EHNR process. Moreover, rotating ring-disk electrode (RRDE) was used to further demonstrate the promoted H* generation for EHNR (Fig. 4d, e and Fig. S20, Supporting Information). In pure PBS electrolyte, the ring current was low due to the limited number of H* on the surface of M-FeSi. With the participation of Ph-NO$_2$, the ring current arose rapidly because of the oxidation of generated H$_2$, which were originated from abundant active H*. Similar phenomenon was then confirmed by the i-t tests. After holding the disk potential of −0.3 V vs. RHE for 10 min, the ring current of PBS with Ph-NO$_2$ system was stabilized at 0.3 mA whereas the ring current of PBS system was negligible.

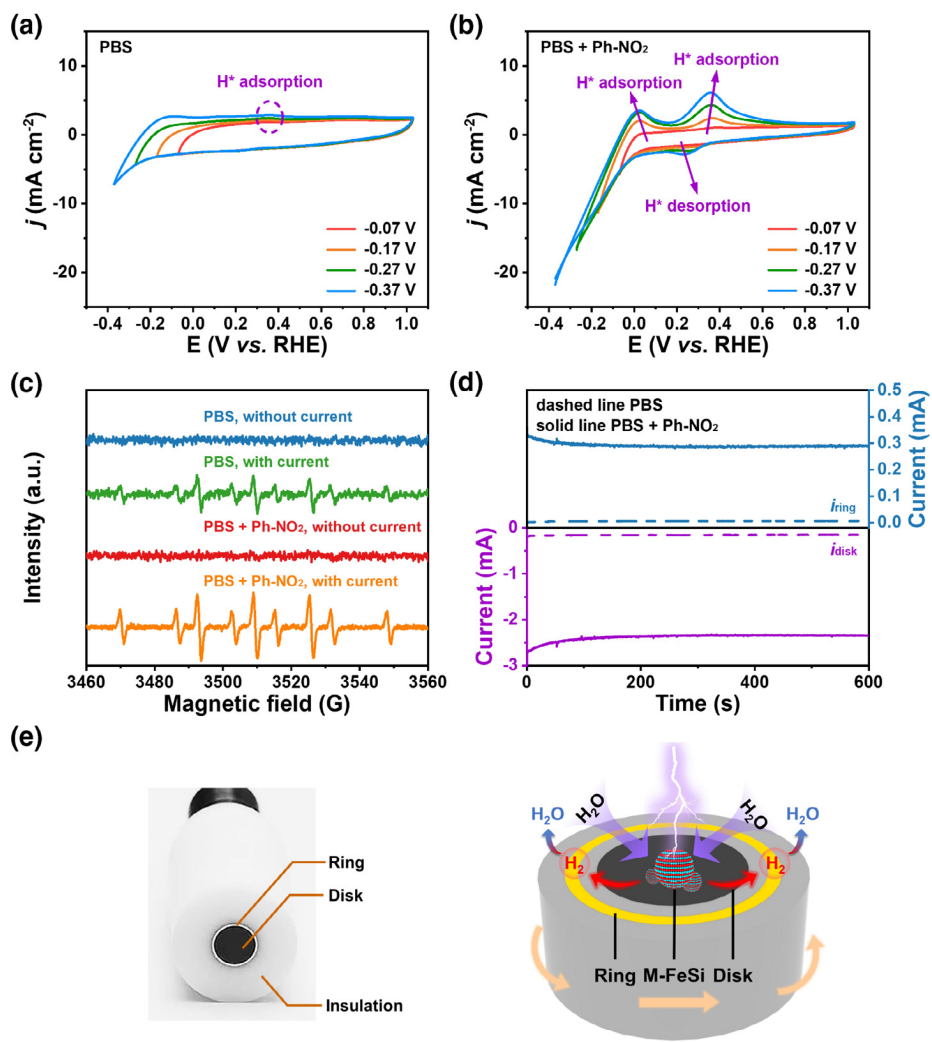

**FIGURE 4**
CV curves of M-FeSi electrocatalyst with a scanning rate of 50 mV s$^{-1}$ under different CV starting potentials from −0.07 to −0.37 V vs. RHE in a) 0.01 M PBS and b) 0.01 M PBS with 10 mM Ph-NO$_2$. c) DMPO spin trapping EPR spectra of M-FeSi electrocatalyst under different system. d) Disk and ring currents in the absence and presence of Ph-NO$_2$ in RRDE test. e) Photo of RRDE and scheme of RRDE operation.

Operando electrochemical impedance spectroscopy (EIS) measurements between 0.2 and −0.4 V vs. RHE were carried out to intensively depict the behaviors of EHNR over M-FeSi. The equivalent circuit for EIS measurements is simplified into two parts (from left to right): 1) resistor–capacitor (R-C) with respect to H* generation, and 2) R-C with respect to EHNR/HER, respectively (Fig. S21, Supporting Information). Compared with EHNR/HER, the H* generation step is associated with peaks at the low frequency (<10$^0$ Hz) [22]. Bode plots in PBS electrolyte without and with Ph-NO$_2$ are shown in Fig. S22 (Supporting Information). For pure PBS electrolyte, the characteristic peaks from 0.2 to −0.1 V vs. RHE corresponded to the accumulation of newly-formed active H*. Staring from −0.2 V vs. RHE, the accumulated active H* intermediates were desorbed from the surface of M-FeSi to release H$_2$. For PBS electrolyte with the presence of Ph-NO$_2$, the characteristic peaks related to H* generation rapidly diminished and the intensity of the characteristic peak related to EHNR increased after 0.0 V vs. RHE, once again indicating that EHNR thermodynamically emerged prior to HER. This phenomenon is consistent with most reported electrocatalytic hydrogenation processes [23]. Meanwhile, surface pH values over M-FeSi were real-time monitored to evaluate the concentration

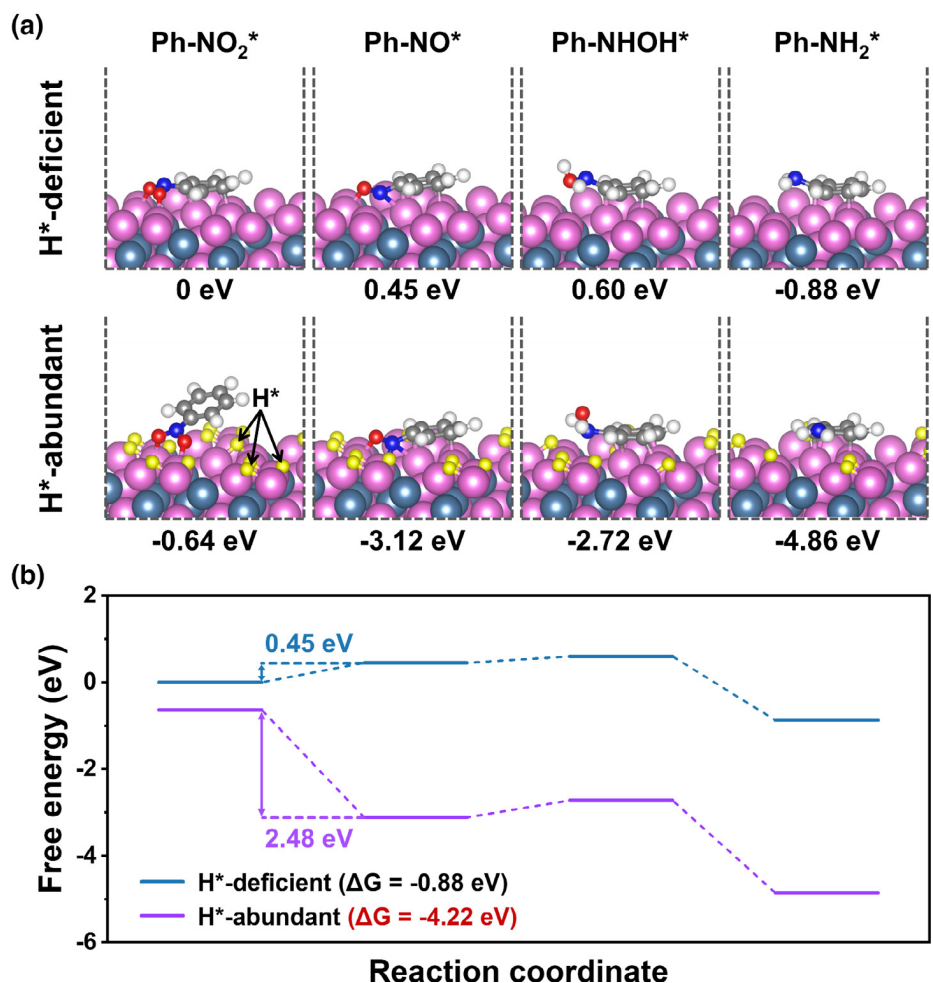

FIGURE 5

a) Calculated Gibbs free energies of Ph-NO$_2$*, Ph-NO*, Ph-NHOH*, Ph-NH$_2$* intermediates, and b) corresponding free energy diagrams on the H*-deficient and H*-abundant surface of M-FeSi.

of H* during the EHNR process (Fig. S23, Supporting Information). Considering that active H* in the neutral medium would be supplied through Volmer step (H$_2$O + e$^-$ → H* + OH$^-$) [24], the rapid increase of surface pH value with the addition of Ph-NO$_2$ demonstrated the high production rate of OH$^-$ from water, which laterally confirmed the fast generation rate of active H* for subsequent EHNR process. Furthermore, to differentiate the contribution of HER and EHNR, the time-dependent charge of these two competitive reactions was quantified by the amount of generated H$_2$. The corresponding gas chromatography (GC) data are illustrated in Figs. S24-S26 (Supporting Information). As summarized in Fig. S27 (Supporting Information), the contribution of HER was negligible in comparison with EHNR, revealing that H* intermediates are preferentially reacted with Ph-NO$_2$ over M-FeSi.

Except for experimental study, the theoretical explanation on the key role of H* during EHNR was detailed illustrated using density functional theory (DFT) calculations. Two models with H*-deficient and H*-abundant M-FeSi surfaces (Fig. 5a and Fig. S28, Supporting Information) were built for comparison. Specifically, for H*-abundant case, the surface of M-FeSi was covered with H* homogeneously, which are represented by yellow spheres. Based on aforementioned proposed reaction pathway, the Gibbs free energies of important steps with the involvements of Ph-NO$_2$*, Ph-NO*, Ph-NHOH*, and Ph-NH$_2$* were calculated and thus the corresponding free energy diagrams were plotted as Fig. 5b. At the H*-deficient surface, the rate-determining step (RDS) was the hydrogenation from Ph-NO$_2$* to Ph-NO* with a Gibbs free energy change ($\Delta G$) of 0.45 eV. Alternatively, at the H*-abundant surface, the RDS was switched to the hydrogenation from Ph-NO* to Ph-NHOH* with $\Delta G$ = 0.40 eV and the first step of Ph-NO$_2$* → Ph-NO* became spontaneous. As a result, the overall $\Delta G$ of EHNR was decreased from −0.88 to −4.22 eV with the aid of ample H* on the surface of M-FeSi, denoting that high concentration of active H* intermediates facilitates the neutral EHNR process toward Ph-NH$_2$ [25].

## Conclusions

A solvent-free one-step microwave methodology was developed to design the noble-metal free topological M-FeSi catalyst for highly efficient EHNR toward Ph-NH$_2$ through the integration of experimental work and theoretical calculation. Compared with A-FeSi (arc melting method), their single counterparts and other candidates, M-FeSi catalyst possesses remarkable conversion of 99.7% Ph-NO$_2$ and 93.8% yield of Ph-NH$_2$ with FE of 83.1% in neutral electrolyte. Besides, M-FeSi catalyst could retain excellent EHNR activity over 21 h without the obvious degradation. The superior electrocatalytic hydrogenation performance is attributed to the accelerated generation of active H* intermediates on the surface of M-FeSi catalyst, which are confirmed by experimental results, including CV, in situ EPR, and RRDE. Furthermore, DFT calculations of H*-deficient and H*-abundant surfaces of M-FeSi denote that the RDS is changed from Ph-NO$_2$* → Ph-NO* to Ph-NO* → Ph-NHOH* with the aid of abundant surface active H* intermediates, which facilitates the conversion to Ph-NH$_2$. This work highlights new insights of EHNR and paves the way for developing high-performance topological TMSi electrocatalyst.

## CRediT authorship contribution statement

**Yuchen Wang:** Conceptualization, Methodology, Formal analysis, Investigation, Writing – original draft, Visualization. **Yaoyu Liu:** Methodology, Validation, Formal analysis, Investigation, Data curation, Writing – original draft, Visualization. **Zhiyue Zhao:** Formal analysis, Investigation. Zhikeng Zheng: Investigation. **Alina M. Balu:** Writing – review & editing, Supervision. **Rafael Luque:** Writing – review & editing, Funding acquisition. **Kai Yan:** Conceptualization, Methodology, Resources, Writing – review & editing, Supervision, Project administration, Funding acquisition.

## Data availability

Data will be made available on request.

## Declaration of Competing Interest

The authors declare that they have no known competing financial interests or personal relationships that could have appeared to influence the work reported in this paper.

## Acknowledgements

This work is supported by National Key R&D Program of China (2020YFC1807600), National Ten Thousand Talent Plan, National Natural Science Foundation of China (22078374), Key-Area Research and Development Program of Guangdong Province (2019B110209003), the Scientific and Technological Planning Project of Guangzhou, China (No. 202206010145) and Hundred Talent Plan (201602) from Sun Yat-sen University, China. R. Luque acknowledges support provided by the Center for Refining & Advanced Chemicals (CRAC) at King Fahd University of Petroleum and Minerals (KFUPM). The support of KFUPM, Dhahran, Saudi Arabia, is also highly appreciated.

## Appendix A. Supplementary material

Supplementary data to this article can be found online at https://doi.org/10.1016/j.mattod.2023.04.016.